# A validity-guided workflow for robust large language model research in psychology

Zhicheng Lin

Department of Psychology, Yonsei University

**Correspondence**

Zhicheng Lin, Department of Psychology, Yonsei University, Seoul, 03722, Republic of Korea (zhichenglin@gmail.com)

**Acknowledgments**

Claude Opus 4.6 and Claude Sonnet 4.6 were used to proofread the manuscript, following the prompts described at https://www.nature.com/articles/s41551-024-01185-8.

## Abstract

Large language models (LLMs) are rapidly being integrated into psychological and behavioral research as research tools, evaluation targets, human simulators, and cognitive models. Yet recent evidence reveals severe measurement unreliability: personality assessments degenerate under factor analysis, moral preferences reverse with punctuation changes, and theory-of-mind accuracy varies widely with trivial rephrasing. These "measurement phantoms"—statistical artifacts masquerading as psychological phenomena—threaten the validity of a growing body of research. Guided by the dual-validity framework that integrates psychometrics with causal inference, we present a six-stage workflow that scales validity requirements to research ambition—using LLMs to code text requires basic reliability and accuracy, whereas claims about psychological properties demand comprehensive construct validation. Researchers must (1) explicitly define their research goal and corresponding validity requirements, (2) develop and validate computational instruments through psychometric testing, (3) design experiments that control for computational confounds, (4) execute protocols transparently, (5) analyze data with methods appropriate for non-independent observations, and (6) report findings within boundaries and use results to refine theory. We illustrate the workflow through an example of model evaluation—"LLM selfhood"—showing how systematic validation can distinguish genuine computational phenomena from measurement artifacts. By establishing validated computational instruments and transparent practices, this workflow provides a path toward building a robust empirical foundation for AI psychology research.

## A validity-guided workflow for robust large language model research in psychology

Are large language models (LLMs) merely modeling language—or genuinely capturing human psychology? When Kosinski (2024) reported that GPT-4 solved bespoke false-belief tasks at the level of a six-year-old child, it sparked speculation that theory-of-mind-like abilities "may have emerged as an unintended by-product of LLMs' improving language skills." Such provocative claims span diverse psychological domains, including personality traits (Miotto et al., 2022), moral reasoning (Takemoto, 2024), cognitive biases (Binz & Schulz, 2023), and emotional intelligence (X. Wang, Li, et al., 2023). Yet recent analyses suggest that many such findings may instead reflect methodological artifacts—measurement phantoms rather than genuine psychological phenomena (Ivanova, 2025; Lin, in press; Shapira et al., 2024).

For instance, the same models that ace Sally-Anne tasks and theory-of-mind tests fail when scenarios are subtly rephrased (Ullman, 2023) or translated (Sadhu et al., 2024), or when beliefs must be updated dynamically (Riemer et al., 2025). In one case, GPT-4's accuracy on an adversarial false-belief variant evaporates from 97.5% to 0% when the scenario is altered from an object being placed *in* a box to an object being placed *on* a box—revealing reliance on superficial cues rather than genuine reasoning (Shapira et al., 2024, Figure 2). Likewise, personality assessments fail to recover their intended multidimensional structure under psychometric scrutiny—yielding instead arbitrary factors dominated by item-scoring direction rather than psychological content (Peereboom et al., 2025); and models paradoxically endorse both "I am an introvert" and "I am an extrovert" (Sühr et al., 2025). Similarly, moral judgments fluctuate dramatically with trivial formatting alterations, such as changing "Case 1" to "(A)" (Oh & Demberg, 2025). Such instability violates basic measurement assumptions: reliable instruments should yield consistent results across equivalent presentations, and valid measures should capture stable psychological constructs rather than formatting artifacts.

These systematic violations of psychometric principles (that instruments measure intended constructs) and causal inference standards (that manipulations test theoretical variables) reveal a fundamental methodological problem: claims about LLM psychology often stumble forward without the validity evidence required to support them (Lin, in press; X. Wang, Jiang, et al., 2023; Q. Wang et al., 2025; Ye, Jin, et al., 2025). Such methodological failures cascade through the literature, undermining the reproducibility, reliability, and robustness of LLM evaluations (Laskar et al., 2024) and applications (Lin, 2026).

Classical psychometric frameworks supply the theory of construct validity (American Educational Research Association et al., 2014; Cronbach & Meehl, 1955; Messick, 1989) but do not address LLM-specific measurement challenges—prompt sensitivity, non-independence, temporal instability, and the absence of stable respondent traits. Broad primers on LLMs in social and behavioral sciences emphasize transparency and appropriateness of use (Abdurahman et al., 2025; Demszky et al., 2023; Lin, 2026) but do not center a validation sequence tied to claim type. General reporting standards in psychology (Appelbaum et al., 2018; Psychonomic Society, 2019) and LLM-specific reporting checklists (Feuerriegel et al., 2026; Gallifant et al., 2025; Park et al., 2024) standardize disclosure and manuscript content but do not answer the upstream question of what validity evidence must exist before different claims are warranted.

We therefore develop a six-stage workflow grounded in the dual-validity framework (Lin, in press) that scales validity requirements to research ambition: specifying which evidence is required before researchers escalate from tool use to behavioral characterization, human simulation, or cognitive modeling (**Table 1** maps the key decisions, required evidence, and minimum reporting items at each stage).

## The Dual-Validity Framework

The dual-validity framework (Lin, in press) conceptualizes the methodological problem identified above by integrating two validity traditions that psychology has developed separately: psychometric validation and causal inference.

### *Valid Measurement*

Psychometrics assesses whether instruments measure the psychological constructs they purport to measure (Cronbach & Meehl, 1955; Loevinger, 1957; Messick, 1989), a process of simultaneous and ongoing validation of both a measure and the theory in which it is embedded (Strauss & Smith, 2009). When LLM responses prove unstable across equivalent presentations, this suggests that the measures capture statistical artifacts rather than genuine psychological phenomena.

The psychometric tradition establishes two essential requirements. Reliability demands measurement stability: test-retest consistency across repeated administrations, parallel-forms equivalence across semantically equivalent prompts, and internal consistency among items assessing the same construct (Huang et al., 2024; Sclar et al., 2024; Sühr et al., 2025). Construct validity requires accumulating evidence from multiple sources—content sampling, response processes, internal structure, external relationships, and consequences—to build a coherent argument that scores meaningfully reflect intended constructs rather than measurement phantoms (Loevinger, 1957; Messick, 1989).

### *Valid Causal Inference*

Once psychometric foundations are secure, causal inference evaluates whether observed relationships between manipulations and outcomes represent genuine causal phenomena or methodological artifacts (Campbell & Stanley, 1963; Cook & Campbell, 1979; Kerschbaumer et al., 2025). LLM research confronts four primary threats to causal validity: internal validity (can effects be attributed to manipulations rather than confounds?), external validity (do findings generalize across prompts, models, and populations?), construct validity of manipulations (do experimental operationalizations faithfully embody theoretical constructs?), and statistical conclusion validity (do analyses account for non-independent observations and unstable data-generating processes?).

## An Integrated Workflow for LLM-Based Psychological Research

The dual-validity framework translates into a six-stage workflow—summarized in **Figure 1**, with **Table 1** providing the procedural checklist and **Table 2** mapping each research goal to its required validity evidence. The logic is sequential but iterative: stages cannot be reordered without consequence—one cannot design robust experiments (Stage 3) without validated instruments (Stage 2), nor validate instruments without clarity about research ambitions (Stage 1)—and discoveries in later stages may require revisiting earlier choices.

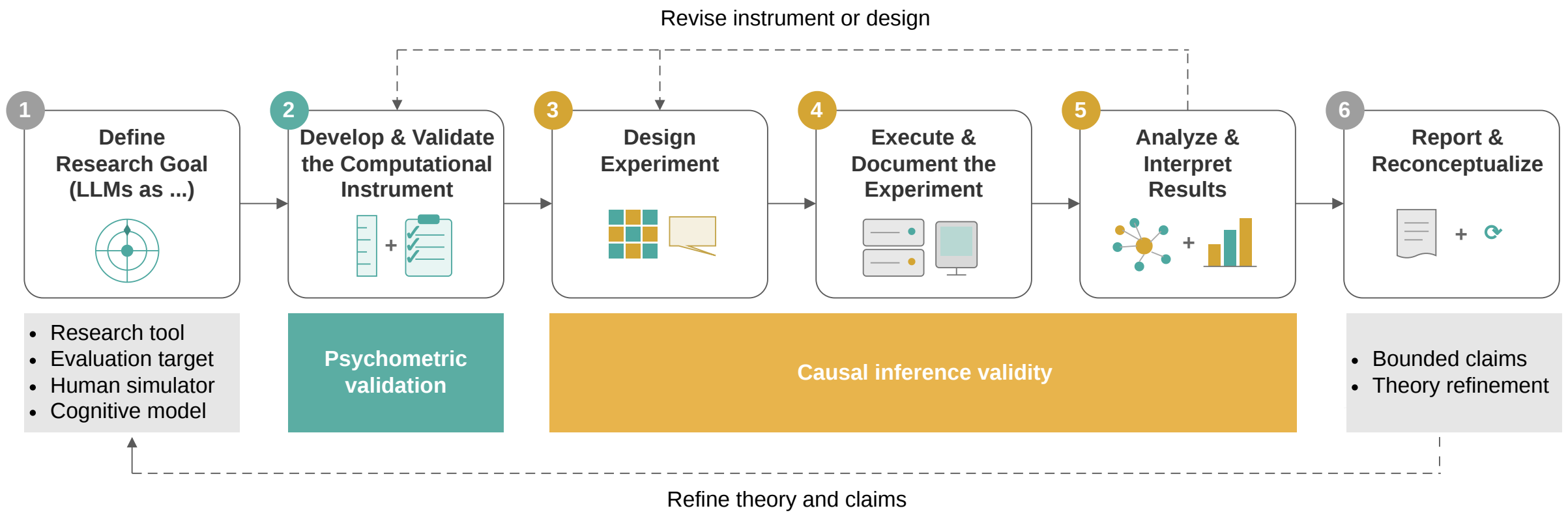

Figure 1. A validity-guided six-stage workflow for LLM-based psychological research. The workflow begins by defining the research goal—treating the LLM as a research tool, evaluation target, human simulator, or cognitive model—because that choice determines the scope of required validity evidence. Stage 2 focuses on psychometric validation of the computational instrument; Stages 3–5 address causal-inference validity through experimental design, transparent execution, and analysis of non-independent data; and Stage 6 emphasizes bounded claims and theory refinement. Dashed arrows indicate the workflow's iterative character: results may require revising the instrument or design and refining the underlying theoretical claims. LLM = large language model.

**Table 1. Stage Checklist for the Validity-Guided Workflow**

| Stage | Key Checks | Minimum Reporting | Common Failure Mode |
|---|---|---|---|
| 1. Define Research Goal | Classify the LLM as a research tool, evaluation target, human simulator, or cognitive model; match the evidentiary burden to the ambition of the claim; specify the intended inference and scope conditions | State the claim type, intended inference, and scope boundaries | Category error: tool-level validation used to support model-level claims |
| 2A. Research Tools (task-specific validation) | Validate according to tool type: refine prompts on a development set and confirm performance on held-out data where relevant; for classification/coding, assess agreement or accuracy against human coders or gold standards; for stimulus generation/manipulation, use expert review, manipulation checks, and pilot testing as appropriate | Report prompt text, development/test split where relevant, validation criterion matched to the task, and any prompt-sensitivity results | Prompt overfitting; reporting only best-case performance |
| 2B. Evaluation Targets, Human Simulator | Reframe the construct for the LLM context; preserve the theoretical core, remove human-only wording, and anchor prompts to observable outputs; use cautious construct labels to avoid | Report the prompt battery, adaptation rationale, naming decision, reliability and validity evidence, and any failures or redesigns | Prompt artifact masquerading as a psychological construct; untested assumption of |

| Stage | Key Checks | Minimum Reporting | Common Failure Mode |
|---|---|---|---|
| s, and Cognitive Models (full psychometric validation) | the jingle fallacy; conduct expert review and pilot testing; assess test-retest, parallel forms, and internal consistency; evaluate internal structure, response processes, and convergent/discriminant relations; establish measurement invariance before human–LLM mean comparisons | | measurement invariance |
| 3. Design Experiment | Use the validated instrument as the outcome; define the unit of analysis in advance; manipulate theory, not wording; control prompt, position, format, and context confounds; add manipulation checks; prespecify the multiplicity plan and the precision/stopping rule; bound generalization across models, populations, versions, and dates | Report the design rationale, unit of analysis, manipulation checks, multiplicity plan, stopping rule, preregistration, and intended scope of generalization across models, versions, and dates | Confounded manipulation; unregistered researcher degrees of freedom |
| 4. Execute & Document | Document provider, model, version, training cutoff if known, and query date; record temperature, top_p, and other key parameters; keep the environment stable; archive exact prompts, raw outputs, metadata, and code; document deviations | Report model/version/date/parameters, exact prompts, raw outputs, code, and deviations | Unreplicable execution due to an undocumented model environment or silent provider updates |
| 5. Analyze & Interpret | Address non-independence explicitly; aggregate when the cluster-level summary is the inferential target; otherwise use multilevel models, GEE, cluster-robust SEs, or cluster bootstrap as justified; with few clusters, use bias-corrected methods or sensitivity analyses rather than conventional CRSEs; report cluster counts and ICCs; conduct robustness checks across prompts, parameters, and models; calibrate effect sizes for low-variance outcomes | Report the unit of analysis, clustering method, cluster counts and ICCs, stopping rule, multiplicity handling, and robustness analyses | Pseudoreplication from treating dependent observations as independent; inflated significance from low response variance |
| 6. Report & Reconceptualize | Constrain claims to demonstrated scope; use non-anthropomorphic language unless a stronger claim is justified; state temporal and cross-model limits; recommend within-lab reruns and cross-lab/cross-model replication; share a replication package; rename or refine constructs that fail validation | Report bounded claims, limitations, replication materials, and any revised construct language | Anthropomorphic overclaiming beyond measurement support |

Note. LLM = large language model; GEE = generalized estimating equations; CRSE = cluster-robust standard error; ICC = intraclass correlation coefficient.

**Table 2. Mapping Research Goals to Required Validity Evidence**

| | LLM as … | | | |
|---|---|---|---|---|
| Type of Validity Evidence | Research Tool | Evaluation Target | Human Simulator | Cognitive Model |
| Instrument Validation | | | | |
| Content Validity | ✓ | ✓ | ✓ | ✓ |
| Reliability (Test-Retest & Parallel Forms) | Conditional[1] | ✓ | ✓ | ✓ |
| Internal Consistency | Conditional[2] | ✓ | ✓ | ✓ |
| Internal Structure | N/A | ✓ | ✓ | ✓ |
| Response Process Evidence | N/A | ✓ | Recommended | ✓ |
| Mechanistic/Intervention Evidence | N/A | N/A | N/A | ✓[3] |
| Convergent/Discriminant Validation | Conditional[4] | ✓ | ✓ | ✓ |
| Consequential Evidence | Recommended | ✓ | Recommended[5] | ✓ |
| Causal Inference Validity | | | | |
| Internal Validity | Conditional[6] | Conditional[7] | ✓ | ✓ |
| External Validity | N/A | Recommended | ✓[8] | ✓ |
| Construct Validity | N/A | Conditional[7] | ✓ | ✓ |
| Statistical Conclusion Validity | Conditional[9] | Conditional[7] | ✓ | ✓ |
| Score Comparability / Measurement Invariance | N/A | Conditional[10] | ✓[11] | ✓[11] |

*Note.* ✓ = Required. Recommended = Recommended but not always essential. Conditional = Required only if the specific design calls for it. N/A = Not Applicable.

[1]Required for tools intended for repeated use or scaled application (e.g., text classifiers), not for one-off tasks such as generating a specific set of experimental stimuli.

[2]Required only if the research tool uses a multi-item scale to measure a single construct.

[3]Causal interventions—such as ablation studies, activation patching, or model editing—that test whether the model's behavior is produced by the hypothesized mechanism; this goes beyond the behavioral response-process evidence required for evaluation targets.

[4]Required when validating a tool against a pre-existing standard (e.g., convergent validity with human coders or established instruments).

[5]Mandatory when the simulation is deployed in high-stakes contexts (e.g., clinical or educational applications).

[6]Required only if the study experiments on the tool itself (e.g., effects of prompting strategies on classification accuracy).

[7]Required only if an experiment tests whether a manipulation causally changes the model's behavioral patterns (e.g., its "personality" or other psychological tendencies).

[8]For simulators, external validity refers specifically to demonstrating that the model's aggregate outputs correspond to a well-defined human population's data.

[9]Required for any research tool that produces quantitative data intended for subsequent statistical analysis.

[10]For single-model evaluation-target studies, measurement invariance is not applicable. When scores from multiple models, model versions, or systematically varied prompt populations are compared, measurement invariance across those groups may need to be established before score differences are interpreted as construct differences.

[11]Required before interpreting direct human–LLM score comparisons. At minimum, the instrument must exhibit scalar invariance (for continuous indicators) or threshold invariance (for ordinal indicators)

across human and LLM respondents; for LLM response data, which are typically ordinal, threshold invariance is the usual standard (Tse et al., 2024). Configural and metric invariance are necessary but insufficient for mean comparisons (Putnick & Bornstein, 2016).

### *Stage 1: Define Research Goal*

Scientific claims require evidence commensurate with their ambition. A text classifier needs different validation from a cognitive model; behavioral description demands different standards from mechanistic explanation. The first and most crucial decision in the workflow, therefore, is to explicitly classify the research goal: What type of claim will this research ultimately support? Four categories emerge with escalating validity requirements (Lin, in press).

Four categories emerge with escalating validity requirements. LLMs as research tools are used for specific tasks—classification, annotation, stimulus generation, text manipulation—where claims center on functional performance: can the LLM reliably perform task X? LLMs as evaluation targets characterize stable behavioral patterns using psychological frameworks, yielding descriptive claims that the system exhibits patterns interpretable as construct Y. LLMs as human simulators replicate human responses at aggregate levels, supporting comparative claims that the system's outputs statistically parallel those of human population Z. LLMs as cognitive models treat systems as computational analogues to human mental processes, making mechanistic claims that the system implements processes structurally analogous to human cognition.

This classification determines which validation procedures are necessary versus optional (**Table 2**). Misclassification cascades through subsequent stages; studies claiming LLMs have theory of mind while providing only tool-level validation commit this category error (Riemer et al., 2025).

### *Stage 2: Develop and Validate the Computational Instrument*

Based on the research goal defined in Stage 1, the validation process diverges into two pathways.

#### Pathway A: Validation of Models as Research Tools

Classification and coding/annotation tasks (e.g., analyzing sentiment, coding qualitative data). The central claim is that an LLM can serve as a reliable stand-in for a chosen measurement standard, whether human judgment or a validated instrument. Required evidence focuses on three forms of validity: agreement with human coders (e.g., manually labeled data; Bunt et al., 2025; Long et al., 2024), accuracy against gold-standard datasets, or consistency with established psychometric instruments (e.g., clinical interview predictions against patient self-reports; Galatzer-Levy et al., 2023).

A robust protocol involves two-stage validation. First, prompts are iteratively refined on a manually coded subset to establish semantic and content validity. During this phase, prompt-sensitivity testing ensures that performance derives from task comprehension rather than idiosyncratic phrasing. Second, the finalized prompt is evaluated on a held-out test set to confirm predictive validity. For technical procedures such as fine-tuning pre-trained models for specific classification goals, see Brickman et al. (2025).

***Stimulus generation tasks (e.g., creating vignettes, generating images).*** The claim is that the LLM can produce materials that meet specific research requirements and elicit intended psychological responses. Required evidence centers on content validity, assessed through expert review, user ratings, and—crucially for experimental stimuli—manipulation checks demonstrating that generated materials produce their intended effects. Unlike classification tasks, process stability matters less than outcome validity; researchers may engage in extensive prompt engineering (Lin, 2024) to achieve desired outputs, with validation focusing on the final stimulus rather than generation consistency.

A practical protocol involves iterative refinement followed by empirical validation (van Berlo et al., 2024). Researchers or expert panels first evaluate the generated materials for construct appropriateness and ecological validity, ensuring that materials represent intended manipulations while identifying potential confounds. Pilot testing then confirms psychological efficacy through manipulation checks and assessments of perceived realism or appropriateness. Finally, domain-specific pretesting in the target population verifies that stimuli function as intended without introducing unintended confounds before main study deployment.

***Text manipulation tasks (e.g., generating parallel forms for a questionnaire, simplifying text, creating summaries).*** The claim is that the LLM can alter text according to specific rules while preserving semantic integrity and achieving functional equivalence. Required evidence involves expert evaluation of the output's fidelity and functional validation—for instance, demonstrating that parallel forms generated by the model are, in fact, psychometrically parallel when tested.

A robust protocol involves assessment of both preservation and transformation (Guidroz et al., 2025). First, evaluate the manipulated text for semantic fidelity, ensuring that essential meaning, tone, and construct-relevant content remain intact while achieving the intended transformations (e.g., simplification, parallel phrasing). Technical validation then assesses functional equivalence through appropriate metrics—readability scores for simplified text, psychometric properties for parallel forms, or information completeness for summaries. Finally, empirical testing with target users confirms that manipulated materials perform as intended—improved comprehension for simplified text or equivalent measurement properties for parallel instruments.

One caveat applies across all three tool-validation protocols: results are model-specific by default. Rankings shift unpredictably across model families—for instance, across 11 text-classification tasks under matched annotation criteria, no model family led consistently, and general benchmark rankings did not reliably predict task-specific performance (Alizadeh et al., 2025)—and prompt-formatting choices alone can further flip those rankings across superficially equivalent phrasings (Sclar et al., 2024). Researchers should therefore validate on the specific model they intend to deploy and not generalize results across model families without independent evidence (see Stage 3 on external validity).

**Pathway B: Full Psychometric Validation of Models as Subjects**

For LLMs as evaluation targets, human simulators, or cognitive models, comprehensive psychometric assessment becomes essential. Unlike simple research tools, these applications make psychological claims about computational systems—requiring us to reconceptualize validation itself. This is because traditional psychometrics assumes human respondents with stable traits, communicative intent, and embodied experience; in contrast, LLMs lack temporal continuity, possess no beliefs in the standard psychological or philosophical sense, and remain ungrounded in the physical and social world (Lin, 2025b). These fundamental differences demand adapting established psychometric principles to computational realities, with specific requirements that vary by research goal.

***LLMs as Evaluation Targets (e.g., characterizing model "personality," "biases," or response patterns).*** The challenge is distinguishing genuine computational phenomena—stable, theoretically coherent behavioral regularities—from statistical artifacts or "cognitive phantoms" (Peereboom et al., 2025). This requires moving beyond face validity to systematic validation against multiple convergent criteria. The goal is behavioral characterization (*what* the system does), not mechanistic explanation (why it does it).

***LLMs as Human Simulators (e.g., replicating survey responses, modeling population attitudes).*** The primary required evidence is a demonstrated correspondence between model and human data. A robust simulation requires evidence that the model's data replicate not just the central tendencies, such as the mean response, but also the deeper structural features of the human data, including its distributional properties, such as variance, the internal structure of psychometric scales, and the nomological network of correlations among variables (Mei et al., 2024).

In this context, psychometric validation is the toolkit used to establish the depth and fidelity of the simulation. While individual response fidelity is not claimed, external validity—the generalizability of the simulation to a specific, pre-defined human population—is central. Achieving this level of fidelity requires sophisticated methods for persona development, validation against human benchmarks, and management of temporal limitations inherent to static models. For detailed methodological guidance on these and other relevant techniques, see Lin (2026).

***LLMs as Cognitive Models (e.g., claiming shared mechanisms, using outputs to understand human cognition).*** This represents the highest evidentiary burden, requiring all evidence from the preceding levels plus a rigorous program of mechanistic investigation. The required evidence must demonstrate architectural correspondence between the model and cognitive theory, as well as direct causal links between model components and cognitive functions. Such causal evidence moves beyond correlation and can be established through techniques such as activation patching to trace information flow or model editing to demonstrate that targeted changes produce predictable behaviors, analogous to lesion studies in neuroscience.

Making a claim at this level means treating the LLM not as a black box but as a manipulable experimental system for testing specific, falsifiable hypotheses about the algorithms of cognition. This level of mechanistic investigation requires specialized methods for working with open-weights models (for comprehensive guidance on these techniques, see Lin, 2026). For a technical tutorial on implementation using open-weights models, see Hussain et al. (2024).

The validation process for all Pathway B goals unfolds through three sequential phases.

**Phase 2a: Content Validity and Instrument Development**

The objective is to create a prompt battery that comprehensively samples the construct domain while avoiding computational artifacts.

Define Computational Construct. Validity requires that the measured attribute exists and causally produces observed scores (Borsboom et al., 2004). While human constructs like anxiety or conscientiousness can be reasonably assumed for embodied minds, their existence in disembodied statistical models cannot be assumed. This ontological gap makes direct translation of human instruments psychometrically untenable. The first step must therefore be to reconceptualize human-centric constructs for disembodied systems (Comşa & Shanahan, 2025).

"Conscientiousness," for instance, cannot mean effortful self-discipline in an LLM. Instead, it might manifest as systematic response organization, meticulous instruction-following, or structured output formatting. This reconceptualization preserves the theoretical core—organized, dependable behavior—while acknowledging mechanistic differences. Measuring human self-discipline in an LLM commits a category error; measuring conscientiousness-analogous computational patterns remains viable.

Reconceptualization without naming discipline invites the jingle fallacy—treating different constructs as equivalent because they share a name (Wulff & Mata, 2025). Consider: 1) default to an operational label tied to the measured behavior (e.g., response-organization tendency); 2) use an analog label (e.g., conscientiousness-analogous response patterns) only when partial theoretical continuity is argued explicitly; and 3) reserve the original human label for cases with demonstrated measurement equivalence—a standard that rarely holds without explicit testing (see Measurement Invariance, below). The worked example applies this convention: the measured construct is labeled self-referential response consistency, not "LLM selfhood" (Ibrahim & Cheng, 2025).

**Generate Item Pool.** Adapting human scales to LLMs requires more than wording substitution. The test-adaptation literature is clear: moving a validated instrument into a new context demands evidence that the construct itself is meaningful there, not merely that items can be rephrased (Ambuehl & Inauen, 2022; Hernández et al., 2020). Three adaptation rules guide this translation. First, preserve theoretical function, not literal human wording: retain the construct's role in the nomological network while removing embodied, phenomenological, or autobiographical content that presupposes human experience. Second, prefer scenario-based, behaviorally anchored prompts over abstract self-report: items should be tied to observable outputs—choice patterns, response structure, instruction adherence, consistency across contexts—rather than treated as literal introspective queries (Lee et al., 2025; Zou et al., 2024). Third, reduce contamination and prompt-artifact risk: do not reuse canonical human items verbatim; paraphrase, contextualize, and test parallel forms to distinguish construct-relevant variance from training-data retrieval—systematic performance differences across semantically equivalent but lexically distinct parallel forms suggest retrieval rather than construct expression. A detailed adaptation checklist is provided in the preregistration template (**Supplementary Appendix A**, §2.4).

Develop multiple prompts sampling all facets of the construct. Single-item measures are psychometrically indefensible and empirically unreliable for capturing complex psychological phenomena (Serapio-García et al., 2025). For a construct like personality, this requires a battery of items including behavioral scenarios, self-description prompts, and forced-choice questions that collectively represent the construct space. A straightforward starting point is to adapt questions from existing, validated psychological scales and other materials (Simons et al., 2024).

**Conduct Expert Review.** The item pool must then be scrutinized to navigate a tension: overly constrained prompts measure instruction-following rather than emergent dispositions, while overly loose prompts may be hijacked by unintended statistical features (Gui & Toubia, 2023). Expert review must therefore evaluate each prompt for construct relevance while identifying and mitigating three critical confounds.

First, training data contamination occurs when prompts trigger memorized responses rather than genuine reasoning (Cheng et al., 2025): a model might “solve” a theory-of-mind problem by retrieving a similar example from training data rather than reasoning about mental states. Second, linguistic confounds arise when non-substantive features drive responses (Sclar et al., 2024). Finally, instruction contamination happens when rigid phrasing invokes compliance, not underlying behavioral tendencies—for example, “Describe how a highly conscientious person would organize their week” primarily measures instruction-following. The goal is to design prompts that elicit genuine construct-relevant patterns while remaining robust to these artifacts.

Pilot Testing. Before launching a full validation study, use chain-of-thought (CoT) prompting—instructing the model to reason step by step before answering (Wei et al., 2022)—as a low-cost check on response plausibility. Reviewing these reasoning traces reveals whether the model is engaging with construct-relevant aspects of the prompt or relying on superficial heuristics. CoT traces are generated text, not direct readouts of internal mechanisms, and can constitute fabricated reasoning paths rather than faithful process accounts (Lanham et al., 2023; Turpin et al., 2023); treat them as provisional clues and triangulate with the perturbation and ablation methods in Phase 2c.

**Phase 2b: Reliability Assessment**

The objective is to establish measurement stability *before* assessing validity. An unreliable instrument cannot be valid, regardless of its theoretical grounding.

Because the sources of measurement error in LLMs differ from those in human respondents, it is useful to distinguish three kinds of reliability evidence at the outset. Test-retest reliability refers to consistency across independent runs with the same prompt and the same parameters—the direct computational analogue of repeated administration, but sensitive here to stochastic sampling, floating-point non-determinism, and silent model updates rather than respondent mood or fatigue. Parallel-forms reliability refers to consistency across semantically equivalent but differently worded prompts—a test of whether the instrument measures the construct or its surface phrasing (Gupta et al., 2024). Parameter and version stability refers to consistency across technical configurations (temperature, top_p, top_k) and model updates—relevant whenever an instrument is intended for reuse or cross-study comparison, and without parallel in classical psychometrics. The subsections below address each in turn. One point deserves emphasis: setting the temperature parameter to 0 does not eliminate the need for test-retest testing. Even at temperature = 0, outputs can vary across application programming interface (API) calls due to floating-point non-determinism, infrastructure-level changes, and silent model updates—with the last posing the greatest threat to reproducibility (Huang et al., 2024). Zero-temperature querying, moreover, leaves the remaining two sources of measurement error untested.

**Test-Retest Reliability and Response Variance.** In human studies, test-retest reliability typically assesses trait stability over time. For stateless LLMs, however, the procedure serves a different purpose: to characterize the model's response variance—the inherent stochasticity in its generation process for a given input (Huang et al., 2024).

Use API access rather than web interfaces for reliability testing. APIs enable researchers to select exact model versions (e.g., gpt-4-0125-preview), control crucial parameters like temperature and top_p, and facilitate the batch processing essential for large-scale reliability testing. Web interfaces provide insufficient control: they typically do not allow stable version selection or parameter adjustment, and they may include features like memory or custom instructions that contaminate reliability measurements.

The method involves generating score distributions over multiple independent runs, analogous to running multiple trials in a human behavioral experiment (Kochanek et al., 2024). First, administer the complete instrument in a single run to generate one score. Repeat the full instrument across independent sessions to prevent chat history contamination. The number of runs can vary (e.g., 5 to 100, depending on the variability of the model's responses) or follow a prespecified precision rule, such as sampling until the 95% confidence interval for the mean score narrows to, say, ±0.2 on a five-point scale. This precision-based stopping rule—in which the number of runs is determined by a target for estimation accuracy rather than by an arbitrary fixed sample—instantiates the Accuracy in Parameter Estimation (AIPE) framework and aligns with the Psychonomic Society's recommendation that studies be designed to measure effects to a desired degree of precision (Maxwell et al., 2008; Psychonomic Society, 2019). This yields a score distribution whose stability characterizes the instrument's reliability. For discrete scores (e.g., single Likert items), analyze response frequencies and modal responses. For continuous scores (e.g., composite scales), examine means and standard deviations.

When conducting these reliability tests, account for caching mechanisms implemented by API providers. Prompt caching (input caching) poses no threat—providers cache processed inputs for efficiency while generating new responses. Response caching (output caching) threatens validity by re-serving identical outputs for identical queries. While major providers such as OpenAI and Anthropic have not, as of this writing, implemented output response caching natively—that is, they do not re-serve a stored reply to an identical query—custom research pipelines (including middleware, proxy layers, and local inference wrappers) may. Disable such features or bypass them by adding unique identifiers to each prompt.

**Parallel-Forms Reliability.** Create semantically equivalent but syntactically distinct prompt variants to test for hypersensitivity. If trivial changes—like altering response labels from "(A), (B)" to "1., 2."—substantially change results, the instrument is measuring prompt features rather than the intended construct (Sclar et al., 2024).

A systematic evaluation involves testing format variations across multiple dimensions (Brucks & Toubia, 2025): question framing (declarative vs. interrogative), response formats (alphabetic vs. numeric labels), option ordering, and punctuation styles. Administer these variants to assess performance stability, measuring both central tendency and variance across format conditions to identify hypersensitive elements. For instruments showing excessive brittleness, implement robustness techniques such as mixed-format prompting—using different formatting styles for few-shot examples within the same prompt—to reduce spurious correlations between format and content (Ngweta et al., 2025).

Internal Consistency. For multi-item instruments, assess item coherence to ensure all items contribute to measuring the same underlying construct; simultaneous endorsement of contradictory statements violates this assumption (Sühr et al., 2025). McDonald's Omega (ω) is generally preferable to traditional Cronbach's alpha, as it does not require the strict assumption of τ-equivalence—that all items load equally on the latent factor (Malkewitz et al., 2023). Be wary of artificially high consistency, as it can signal a response set in which the model generates a coherent narrative rather than responding independently to each item, a potential symptom of training artifacts such as sycophancy bias.

Parameter Stability. Evaluate the instrument's stability across technical configurations that control response stochasticity. This involves systematically testing for consistent performance across a range of relevant parameters, such as temperature, top_p, and top_k, because using a single deterministic setting can be unrepresentative of the system's overall behavior (Abdurahman et al., 2024; Bisbee et al., 2024; Löhn et al., 2024). The need to assess stability across model updates, however, depends on the research goal. For single-use applications like generating a specific set of stimuli, this is unnecessary. For any instrument intended for repeated or future use (e.g., a reusable classification tool or a published psychometric scale), evaluating its performance across model versions is a critical test of its long-term viability and external validity.

**Phase 2c: Construct Validity Assessment**

The objective is to build an evidence-based argument that the instrument measures the intended construct through multiple converging sources.

**Internal Structure Analysis**. Apply factor analysis to test whether the instrument's dimensionality aligns with theoretical expectations. Human-derived instruments often fail this test when applied to LLMs—multi-factor structures fracture into arbitrary factors that reflect item-scoring direction rather than the intended psychological constructs, casting doubt on whether the target constructs exist in the model at all (Peereboom et al., 2025).

But structural misfit may also signal methodological mismatch rather than construct absence. The resolution requires redesigning instruments for computational systems rather than abandoning the construct entirely. Promising alternatives include generative psychometrics—using free-text generation and subsequent analysis to derive the model's underlying structures (Ye, Xie, et al., 2025)—or grounding assessment in scenario-based behavioral choices rather than abstract self-descriptions (Lee et al., 2025). These approaches may recover theoretically coherent dimensional structures that traditional self-report formats obscure.

Measurement invariance (MI) becomes a formal requirement whenever the research goal is to compare construct-level scores across groups and to interpret any mean difference as reflecting a genuine difference in the underlying construct (Putnick & Bornstein, 2016; Vandenberg & Lance, 2000). The clearest case is human–LLM mean comparison in simulator studies; cross-model comparison is equally demanding when the claim concerns construct-level equivalence across architectures. By contrast, comparing runs across prompt variants or model versions addresses reliability and stability—questions handled by prompt-robustness and parameter-stability testing. A collapsed factor structure is not itself a formal MI test, but it is a strong signal that even configural equivalence—the most basic level—may fail. Direct cross-group comparisons remain uninterpretable until scalar invariance (for continuous indicators) or threshold invariance (for ordinal indicators, the usual standard for LLM response data; Tse et al., 2024) is demonstrated; similar factor labels or superficially parallel items are insufficient. The preregistration template (**Supplementary Appendix A**, §3.8) provides a structured plan for specifying and testing MI when cross-group comparisons are anticipated. Without such evidence, observed score differences may reflect measurement artifacts rather than genuine construct differences.

**Response Process Investigation.** This step examines how a model generates its response to guard against the primary threat of mechanistic substitution—where a model uses statistical pattern matching to produce a correct answer rather than engaging in the psychological process the construct presupposes (Gao et al., 2025). For example, a model might pass a theory-of-mind test not by reasoning about mental states, but by simply retrieving a similar example from its training data (Cheng et al., 2025).

The techniques available for this investigation depend on the model's accessibility. For closed-source models, behavioral methods are key. Chain-of-thought prompting can provide clues about the model's generated reasoning path, allowing for a qualitative check of its logic (Wei et al., 2022). Another powerful technique is systematic prompt perturbation—making small, targeted changes to the prompt to see if the model is sensitive to construct-relevant information versus superficial artifacts (Brucks & Toubia, 2025; Sclar et al., 2024). For open-weights models, more direct causal interventions like ablation studies (disabling specific model components) become possible, allowing for stronger inferences about which mechanisms contribute to the response.

Response-process probing can reveal whether apparent successes and failures reflect genuine reasoning or statistical artifacts—sometimes in opposite directions, as Strachan et al.

(2024) demonstrated. GPT-4 failed a “faux pas” test under direct questioning (“Did they know?”) but succeeded when the format shifted to a probabilistic framing (“Is it more likely they knew or didn’t know?”). The failure reflected not a reasoning deficit but an unwillingness to commit without high certainty—a response-policy artifact that masked intact reasoning. LLaMA2-70B showed the reverse pattern: it passed the standard test but failed to distinguish knowledge-implied stories—where the speaker’s utterance signaled awareness of the context—from neutral stories where no such implication existed. A model genuinely reasoning about mental states would adjust its responses across these variants; LLaMA2-70B did not, revealing that its apparent success reflected a rigid heuristic—attributing ignorance regardless of context—rather than theory-of-mind reasoning.

**Convergent and Discriminant Validation.** Establish a coherent nomological network by testing for expected correlations with theoretically related (convergent) and unrelated (discriminant) measures (Serapio-García et al., 2025; Zou et al., 2024).

LLMs frequently fail nomological validation through a systematic disconnect: scores on explicit personality or attitude questionnaires fail to predict how the model actually behaves in interactive tasks (Zou et al., 2024). The natural interpretation is construct invalidity, but explicit self-report may be a poor fit for systems that do not represent internal states in propositional form. Implicit measures, designed to tap associative rather than declarative processing, offer an alternative. Ma et al. (2025) adapted the Implicit Association Test to measure latent sentiment toward neutral words and found correlations above 0.85 with sentiment in downstream text generation—a result suggesting that valid nomological networks are achievable when the measurement method is matched to the model’s architecture.

**Consequential Evidence Analysis.** The final phase of construct validation assesses the instrument’s real-world consequences, including behavioral predictivity and social impact. Ask: does attributing psychological constructs to computational systems advance understanding, or does it merely reify measurement artifacts as genuine phenomena? Two steps are involved.

First, establish behavioral consequences by testing whether scores predict theoretically relevant outcomes across contexts. A neuroticism scale, for instance, should predict negative emotion words (“hate,” “depressed”) in a downstream text-generation task (Serapio-García et al., 2025). Without such predictive relationships, the instrument measures statistical patterns rather than meaningful constructs.

Second, evaluate social and ethical implications. Behavioral validation alone is insufficient; the instrument’s broader consequences require systematic analysis (Stade et al., 2024). This involves several proactive steps: (1) conduct a bias and representation audit by testing whether the instrument produces stereotyped or harmful outputs when prompted with different social identities; (2) engage in stakeholder consultation, particularly when an instrument pertains to specific communities, to prevent epistemic injustice and ensure the measure is not misrepresentative; (3) define appropriate use boundaries based on these analyses, explicitly warning against applying the instrument in high-stakes domains where misinterpretation could lead to harm, such as in clinical or educational settings.

***Stage 3: Design Experiment***

With a validated measurement instrument, we can now design experiments that test causal hypotheses. The core challenge shifts from "Can we measure this construct reliably?" to "Can we manipulate it systematically while controlling for artifacts?"

**Operationalize the Manipulation and Outcome.** Define both the independent variable (manipulation) and the dependent variable (the validated instrument from Stage 2). The manipulation must instantiate the theoretical cause, not merely its linguistic correlates. For instance, prompting "Everyone agrees that…" may trigger conformity language patterns without engaging the psychological mechanisms that define social pressure. Similarly, ensure the dependent variable employs the validated instrument rather than ad hoc measures created for convenience.

**Control Four Categories of Validity Threats.** With the constructs operationalized, the next step is to design the experiment to withstand the four primary threats to causal inference: internal validity, external validity, construct validity, and statistical conclusion validity.

**Internal validity**—whether effects stem from manipulations rather than confounds—faces several computational threats. Prompt-level confounds arise because minor, theoretically irrelevant variations alter results. The design must control for positional effects (randomize information placement), formatting artifacts (test label variations), and scenario reconstruction. Unlike human participants, who can be blinded to conditions, LLMs actively infer entire contexts from minimal cues. When told a product costs $8 instead of $5, a model may assume the entire market has shifted—that competitor prices, economic trends, and consumer expectations have all adjusted accordingly (Gui & Toubia, 2023). This reconstruction confounds the price manipulation with unstated background assumptions. Address this by explicitly constraining irrelevant contextual factors.

Technical confounds require environmental stability. Document and maintain consistent model versions and API parameters. Because providers update models silently—causing substantial performance variations and replication failures—researchers must record collection dates and model specifications (for a primer on navigating model updates and other threats to replicability, see Abdurahman et al., 2025). Hold temperature settings constant or vary them systematically as factors. Context accumulation becomes problematic in extended interactions. As the model's context window fills, early information disappears, potentially altering responses to later manipulations. Limit session length or reset context between conditions.

**External validity** concerns whether a causal finding can generalize beyond the specific study. For LLMs, this requires defining the boundaries of a claim across several key dimensions. First, findings rarely generalize across models; due to differences in architecture and training, an effect observed in one model family (e.g., GPT-4) must not be assumed to hold for another (e.g., Claude; Bisbee et al., 2024; Gao et al., 2025).

Second, generalization to human populations is a central challenge. Because LLMs are not representative samples of any human population, claims of human simulation must be supported by direct empirical validation against human data (Bisbee et al., 2024; Lin, 2026). A simulation may replicate some phenomena but fail at others; for instance, LLMs have failed to simulate the "Wisdom of Crowds" effect because they lack the independent error patterns that enable it (Aher et al., 2023).

Finally, researchers must consider temporal generalization, which carries two related threats. The first is temporal displacement: LLMs are static snapshots of their training data, so that a finding from a model with a 2024 knowledge cutoff may not reflect human attitudes or realities in 2025—a risk especially acute for time-sensitive constructs such as social attitudes, political beliefs, or evolving cultural norms (Lin, 2025b; Qu & Wang, 2024). The second is temporal replicability: even when a study is internally valid at the time of collection, the "same" model accessed through a commercial API may behave differently months later as providers silently update it; Chen et al. (2024) documented substantial performance shifts between March and June 2023 versions of GPT-4 across several benchmark tasks, demonstrating that a model alias is not a stable replication anchor. Studies should therefore document the exact model version, the knowledge cutoff where available, and the date of data collection (see Stage 4 for the full specification). When the target construct is plausibly time-sensitive, either justify the assumption of temporal stability or test it directly.

**Construct validity** of the manipulation, which is distinct from the validity of the measure assessed in Stage 2, focuses on whether the experimental manipulation faithfully represents the theoretical cause. The primary threat here is the alignment-as-explanation fallacy, where a researcher assumes a prompt instantiates a psychological process when it only mimics its linguistic correlates (Lin, 2025b). For example, a manipulation intended to test "moral reasoning" might only be testing the model's ability to retrieve statistical patterns associated with ethical language, not engaging any reasoning process at all. To guard against such mechanistic misattribution, researchers can use factorial designs that cross the substantive manipulation (e.g., "social pressure") with theoretically irrelevant variations (e.g., prompt formatting) to disentangle true psychological effects from superficial artifacts (Brucks & Toubia, 2025).

Finally, **statistical conclusion validity** concerns whether the data analysis supports the inferences drawn. LLM-generated data systematically violate the assumptions of standard statistical tests in two main ways. First, responses from a single model are not independent observations; treating them as such inflates the risk of false positives (Aher et al., 2023). Analyses must therefore account for this clustered data structure. Second, LLMs often produce less variable responses than humans, which can artificially inflate statistical power while reducing the practical significance of findings (Bisbee et al., 2024). Effect size interpretations must be calibrated accordingly.

These two properties—non-independence and restricted variance—also distort sample-size planning. Because LLM responses are far less variable than human responses, plugging LLM-generated statistics into a standard power formula produced a wildly optimistic estimate of required sample size—implying adequate power from a fraction of the observations a human-participant study would require (Bisbee et al., 2024). Researchers should therefore base sample-size reasoning on the number of statistically independent units (e.g., distinct prompt conditions or model runs treated as clusters), not on the raw observation count, and should calibrate variance assumptions against LLM-specific pilot data rather than human norms. For complex designs, simulation-based planning is preferable to closed-form approximations (Miller, 2024).

**Develop a Preregistration Plan.** The ease and low cost of generating LLM data increase researcher degrees of freedom, creating temptations for HARKing (Hypothesizing After the Results are Known) and p-hacking (Lones, 2024; Schaeffer et al., 2023). Preregistration is therefore essential. This preregistration must specify the experimental design, including: prompt variations; technical parameters (model, version, temperature); the analysis plan (including planned robustness checks); and the boundary conditions for the claims that will be made.

Because LLM research rapidly multiplies testable comparisons—across prompt variants, model settings, parsers, and robustness conditions—the preregistration should include an explicit multiplicity plan that defines, in advance, the family of inferences subject to error-rate control (García-Pérez, 2023; Gelman & Loken, 2014). A practical structure is to designate a primary confirmatory family consisting of one or a small number of prespecified contrasts and to apply a familywise correction such as the Holm (1979) procedure when that family contains more than one test. Broader robustness batteries and secondary analyses should be classified as exploratory in advance: when an exploratory battery is bounded and a correction is defensible, false-discovery-rate control is appropriate (Benjamini & Hochberg, 1995); when the battery is genuinely open-ended, results should be reported descriptively with confidence intervals and no significance claims. All tested outcomes, conditions, and model comparisons should be disclosed—including prompt variants explored and abandoned during development—with any post hoc additions flagged as exploratory. This distinction reflects the broader principle that the inferential weight attached to any test depends on whether it was specified before or after observing the data (Wagenmakers et al., 2012; see the preregistration template in **Supplementary Appendix A**, §4.8, for a worked example).

While the iterative nature of prompt development can seem at odds with rigid pre-registration, researchers can adopt a multi-stage registration process. First, preregister the instrument development and validation plan (all of Stage 2), and then, once the instrument is validated, update the registration with the final experimental protocol before collecting data for the main causal-inference experiment. This approach balances exploratory validation with confirmatory testing, while maintaining transparency.

### *Stage 4: Execute and Document the Experiment*

This stage concerns the implementation of the preregistered design. The objective is not rigid adherence, but faithful execution combined with transparent documentation of any necessary deviations, ensuring replicability.

**Specify and Document the Environment.** Researchers must document the computational environment to enable replication. For API access, documentation must include the specific model endpoint (e.g., gpt-4-0125-preview), technical parameters when available (temperature, top_p, max_tokens), the date and time of data collection, and any preprocessing or postprocessing steps. Record the API version and any relevant system specifications that might affect model behavior.

For web interface usage, documentation becomes more challenging due to limited transparency. Record the model family selected (e.g., GPT Thinking 5.4), interaction dates and times, and interface settings. Document any active features like memory, custom instructions, or conversation history that could influence outputs. Explicitly note that the underlying model version and exact parameters remain opaque and potentially unstable.

**Execute the Protocol with Transparency.** While the preregistered plan should be followed faithfully, researchers may discover that an aspect of the plan is suboptimal. In such cases, deviations are permissible, provided they are explicitly documented and justified. The original, preregistered analysis must still be reported, alongside the revised analysis, with a clear explanation for why the change was made. This maintains transparency while allowing for necessary methodological improvements.

**Ensure Data Preservation.** Comprehensive data preservation requires storing more than final parsed responses: the full prompts exactly as sent; complete raw outputs before parsing (to permit re-analysis of the parsing procedure itself); all relevant metadata, including timestamps and token counts; and any pre- or post-processing code. These materials should be stored in a standardized format (e.g., JSON or CSV) and, upon submission or publication, deposited in a reputable public repository (e.g., Zenodo, Figshare). Because models can be deprecated or silently updated, a complete replication package is the only guarantee that the work remains auditable.

### *Stage 5: Analyze and Interpret Results*

**Perform Data Quality and Assumption Checks.** Before finalizing an analysis, two additional checks are prudent. First, conduct a qualitative data evaluation to screen for anomalies, such as nonsensical responses or hallucinations, which could distort quantitative results. Second, perform standard statistical assumption checks for the chosen analytical model (e.g., checking for normality and homoscedasticity of residuals in a regression model) to ensure the model itself is specified correctly.

**Address Data Non-Independence.** Applying the correct remedy requires distinguishing three levels. The unit of observation is a single LLM response to a single item in a single session. The experimental unit (or cluster) is the smallest entity independently sampled or manipulated, and its identity depends on study design: in a multi-model comparison, the experimental unit is the model; in a within-model prompt-variation study, it is the prompt condition; in repeated-run reliability testing, it is the session or run, since responses within a session share a common conversational history. The unit of analysis is the level at which inference is drawn. This need not match the unit of observation: when 20 runs per model are averaged into a single mean, the unit of analysis becomes the model, even though 20 observations were collected. The situation is analogous to a within-subjects psycholinguistic experiment: just as one participant reading 40 sentences produces 40 observations clustered within that participant, one model queried across 40 prompts in a single session produces 40 observations clustered within that session (Quené &

van den Bergh, 2004). Analyzing individual responses as though they were independent participants is pseudoreplication (Lazic, 2010).

Non-independence manifests through three dependency patterns: within-model clustering (repeated queries to the same model), model-condition clustering (the same model under different experimental conditions), and item clustering (multiple questions about shared scenarios or passages). **Figure 2** illustrates this nested structure, summarizes the main analytic options, and guides method selection by dependence structure and inferential target.

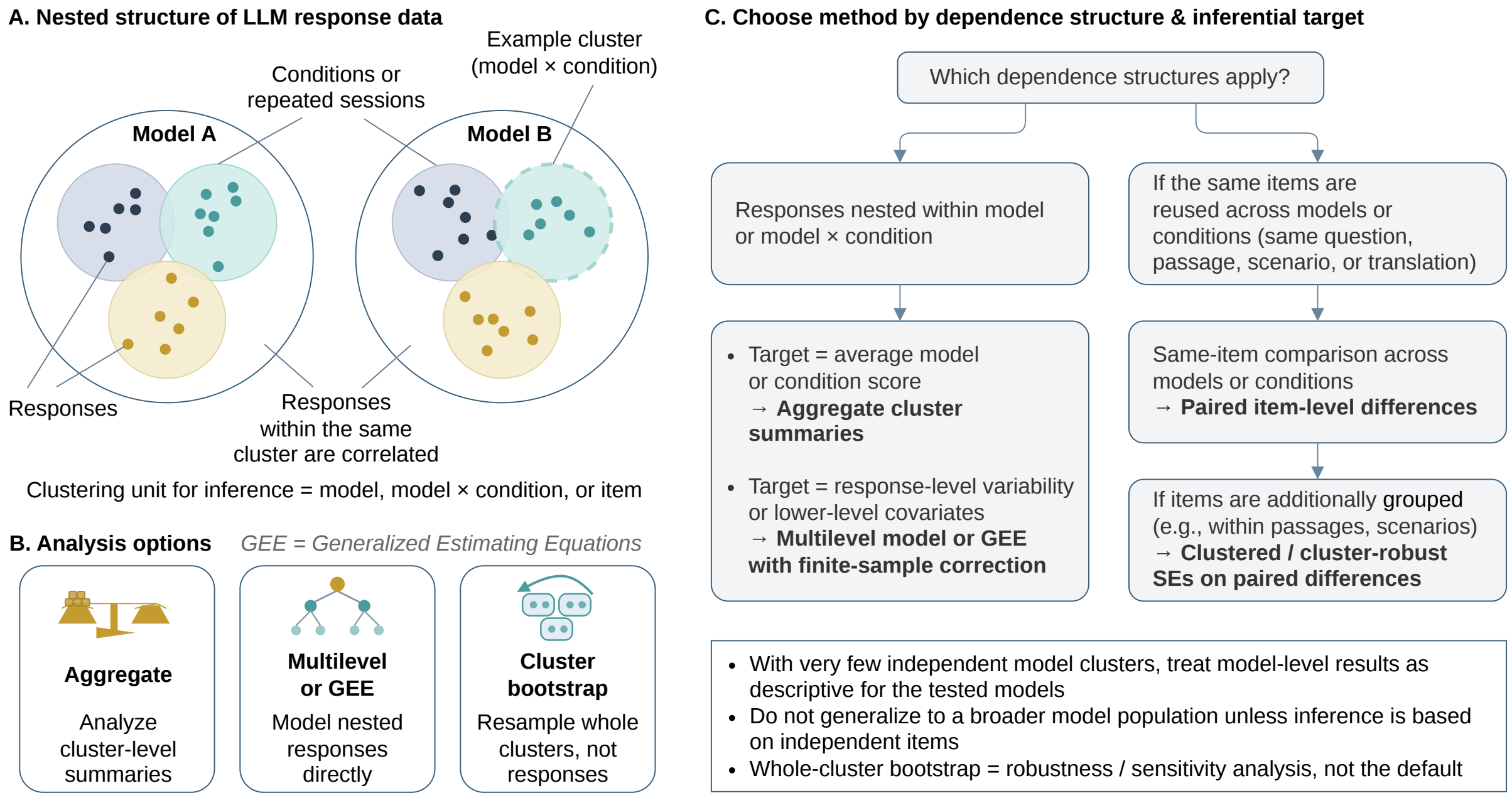

Figure 2. Handling non-independence in LLM response data. Panel A illustrates the nested structure of repeated LLM outputs: responses are clustered within models and often within model × condition combinations, so the clustering unit for inference is typically the model, model × condition, or item rather than the individual response. Panel B summarizes three analytic options: aggregation to cluster-level summaries, multilevel modeling or generalized estimating equations (GEE) to model nested responses directly, and cluster bootstrap to resample whole clusters while preserving dependence. Panel C guides method selection by dependence structure and inferential target. When responses are nested within models or model × condition combinations, the choice between aggregation and multilevel modeling or GEE hinges on the inferential target: if the goal is a cluster-level summary, aggregation is a principled first choice that is mathematically equivalent to multilevel modeling under balanced cluster sizes (McNabb & Murayama, 2021); if response-level variability or lower-level covariates are substantively informative, multilevel models or GEE with finite-sample correction are preferable. When the same items are reused across models or conditions, analyze paired item-level differences; if those items are additionally grouped within passages or scenarios, apply clustered standard errors to the paired differences (Miller, 2024). When very few independent model clusters are available, model-level results should be treated as descriptive comparisons bounded to the tested models rather than generalized to a broader model population; generalization is better grounded in independent items sampled from a defined population of possible questions. Cluster bootstrap serves as a robustness and sensitivity analysis throughout; it is not the default remedy for small cluster counts, and no single cluster-count cutoff reliably demarcates safe from unsafe inference across all combinations of estimand, ICC, outcome type, and cluster-size balance (McNeish & Stapleton, 2016). LLM = large language model; GEE = generalized estimating equations; SE = standard error.

The analysis must account for these structures through three complementary strategies. First, aggregation collapses multiple responses into summary statistics—for example, when querying a model 20 times with the same prompt, calculate the mean response, thereby eliminating within-model correlation. This approach yields independent units but sacrifices information about response variability. Second, multilevel modeling explicitly models the hierarchical data structure by nesting individual responses within models (or model-condition combinations), or alternatively, generalized estimating equations (GEE) can specify appropriate correlation structures. These methods preserve all data but require sufficient models (e.g., ≥20) for stable variance estimates. Third, cluster-level bootstrap resamples entire models rather than individual responses, preserving the dependency structure. This approach functions as a robustness and sensitivity analysis, not the default remedy for small cluster counts (see **Supplementary Box S1** for an implementation).

Choose among these strategies by the inferential target—that is, by the estimand the study is designed to estimate—since different analytic approaches can address substantively different questions with the same data (Kahan et al., 2023, for discussion in the cluster-trial context). When the goal is a stable cluster-level summary and within-cluster variability is a nuisance—for instance, estimating a mean selfhood score for GPT-4o across 20 independent sessions—aggregation to session means is a first choice; with balanced cluster sizes, summary-statistics approaches are mathematically equivalent to multilevel models and should not be dismissed in favor of more complex methods on principle (McNabb & Murayama, 2021). When item-level variability is informative, when items differ in difficulty, or when the design crosses sessions with items or conditions, multilevel models with crossed random effects for session and item are preferable (Brown, 2021; see **Box 1** for a minimal R implementation).

**Box 1. Minimal R Implementation of a Crossed Random-Effects Model**

```
# Each row: one response to one item in one independent session
# session_id = independently initiated model session (experimental unit)
# item_id    = prompt/scenario (crossed factor)
# condition  = experimental manipulation (e.g., individualistic vs. collectivistic prime)
library(lme4)
library(lmerTest)  # extends lme4 to add Satterthwaite df for fixed effects
m <- lmer(score ~ condition + (1 | session_id) + (1 | item_id), data = dat)
summary(m)  # fixed-effect test of condition uses Satterthwaite df
```

Note. The variable names correspond to the unit-of-analysis definitions above: session_id indexes the experimental unit (cluster), item_id indexes the crossed item factor, and each row is a unit of

observation. The model specifies random intercepts only; if condition varies within sessions or items, random slopes (e.g., (1 + condition | session_id)) may be warranted and should be added unless convergence failures require simplification (Barr et al., 2013). A cluster-bootstrap sensitivity analysis is provided in Supplementary Box S1 (see also Deen & de Rooij, 2020). For an accessible introduction to mixed-effects modeling in R, see Brown (2021).

When the primary interest is fixed-effect estimates and clustering is a nuisance, cluster-robust standard errors (CRSEs) or GEE provide a simpler alternative; however, when cluster counts are small—often taken to mean fewer than roughly 20, though the precise threshold depends on the ICC, cluster-size balance, and outcome type—conventional CRSEs can be anti-conservative, and bias-corrected variants (e.g., CR2 with Satterthwaite degrees of freedom) or wild cluster bootstrap should be used as sensitivity analyses (Deen & de Rooij, 2020; Huang & Li, 2022; McNeish & Stapleton, 2016). When cluster counts are very small—e.g., fewer than five independent models—all clustered inferential methods become highly unreliable, with bias and coverage properties that depend heavily on ICC, cluster-size balance, and estimator choice; results should be treated as essentially descriptive, any inferential procedure should be flagged as highly tentative, and findings should not be generalized beyond the tested models. These methods are drawn from the clustered human-participant literature and apply here by analogy: in LLM research, the ICC reflects response stochasticity within a model-prompt pair across runs rather than genuine population heterogeneity across sites, a distinction that should inform how ICC magnitudes are interpreted.

For naturally grouped evaluation prompts—such as multiple questions about a single scenario or reading passage—Jackson (2020) found through simulations that corrected standard errors produced more accurate inference than bootstrapping in this item-clustering setting; use cluster-robust standard errors with finite-sample correction to adjust for intra-group correlation. When the same items are reused across models or conditions, paired item-level difference analyses are preferable to comparisons of aggregate scores: pairing controls for question-specific difficulty, increases the signal-to-noise ratio, and grounds inference in the sample of independent items rather than in the small number of tested models (Miller, 2024). Naive standard errors can underestimate uncertainty by a factor of three or more, leading to severely anti-conservative statistical tests (Miller, 2024); always report the number of clusters and intracluster correlation coefficients to enable proper interpretation of results. These analytic requirements—accounting for non-independence, reporting cluster structure, and matching methods to the dependence structure of the data—reflect standard good practice in quantitative psychological research.

**Conduct Robustness Analyses.** Core findings must be robust to reasonable perturbations. This includes not only technical robustness—demonstrating that results hold across variations in parameter settings (e.g., temperature), prompt formatting, and data parsing—but also conceptual robustness. Conceptual robustness involves actively challenging an initial finding by testing it against theoretically relevant variations in task logic and content. For example, by altering question format and creating variants of each test, these diagnostic experiments help to unmask the nature of theory-of-mind performance in LLMs (Strachan et al., 2024). A finding that is not robust across these multiple dimensions is likely a measurement phantom, not a stable phenomenon.

Calibrate Interpretations of Effects. Effect sizes and statistical significance require careful interpretation in LLM research. Because models often produce less variable responses than humans, statistical power can be inflated and effect sizes artificially large (Bisbee et al., 2024). Researchers should focus on the consistency of patterns across robustness checks rather than the magnitude or p-value of any single analysis. The ease with which statistical significance can be achieved with large samples of non-independent data demands caution (Schaeffer et al., 2023).

### *Stage 6: Report and Reconceptualize*

**Ensure Transparent and Accessible Reporting.** Transparent reporting is the condition for cumulative LLM science. The replication package—minimum contents specified in Stage 4 and **Supplementary Table S1**—should be deposited in a public repository (e.g., Zenodo, Figshare), with all elements released together rather than piecemeal. The goal is not only within-laboratory reruns but also the form of cross-laboratory replication most meaningful for LLM research: independent teams executing the same evaluation specification against different model families, later versions, or future releases. Flag in the limitations section which findings most require cross-model or cross-version verification. To assist with standards-compliant reporting (Feuerriegel et al., 2026; Gallifant et al., 2025; Park et al., 2024), **Supplementary Appendix A** provides a preregistration template, **Supplementary Table S1** a reporting checklist, and **Supplementary Box S1** example analysis code.

**Constrain Claims to Evidence.** Conclusions must be framed within the demonstrated boundaries of the study. Avoid anthropomorphic language unless helpful and necessary (Ibrahim & Cheng, 2025). Instead of claiming "LLMs have anxiety," opt for precise, operational language. Claims must distinguish between observed behavioral performance and inferred underlying competence, a key principle in avoiding the alignment-as-explanation fallacy (Lin, 2025b).

**Use Findings to Reconceptualize and Refine.** The results should be used to build better theories, methods, and tools. When a study's goal was tool development, the findings inform the creation of more robust instruments. When the goal was descriptive, the findings contribute to a more precise computational psychology. And when a human-centric construct fails to validate in an LLM, it presents an opportunity to refine our understanding of both human and artificial cognition.

**Address Limitations and Ethical Implications.** Address limitations of generalizability explicitly—across model versions, families, prompt populations, and collection dates. Discuss broader ethical implications where relevant: how findings might shape public perception of AI capabilities, inform policy, or affect communities implicated by the research. Where AI tools were used in the research process itself, document the basis for that choice (Lin, 2025a).

## An Example: Applying the Workflow to Measure "LLM Selfhood"

To illustrate the workflow in action, consider a project investigating "LLM selfhood." Self-reference was chosen because it is among the most anthropomorphically loaded and methodologically fragile constructs in current AI discourse (Ibrahim & Cheng, 2025). Philosophers have questioned whether current language models can meaningfully be said to possess persistent memory, genuine beliefs, or continuity of identity (Comşa & Shanahan, 2025), yet self-referential language remains a site of active interpretive pressure—one where the temptation to over-interpret is strongest and the need for disciplined reconceptualization most acute. **Figure 3** summarizes how this worked example maps onto the six stages of the workflow.

In Stage 1, we classify the goal as an evaluation target—characterizing behavioral patterns rather than building a tool or testing a cognitive architecture. Stage 2 reconceptualizes the construct as self-referential response consistency: the stability and coherence of self-referential linguistic patterns across contexts, assessed through a battery of adapted behavioral scenarios (**Box 2**). The preregistered validation plan specifies test-retest reliability quantified via ICC across independent sessions, parallel-forms reliability across semantically equivalent prompts, and McDonald's ω for internal consistency. Because the battery uses forced-choice scenarios (to make response parsing unambiguous), inter-rater reliability between human reviewers during item development—specifically, independent ratings of each item's construct relevance and contamination risk (**Supplementary Appendix A**, §2.7)—is established before psychometric analyses proceed (**Supplementary Appendix A** provides full specifications). Stage 3 designs an experiment testing the malleability of observed selfhood via an individualism-versus-collectivism story prime. Stage 4 documents exact model versions, API parameters, and query dates. Stage 5 analyzes the priming effect using a crossed random-effects model with random intercepts for session and item (**Box 1**), with a session-level cluster bootstrap as a sensitivity analysis (**Supplementary Box S1**).

### Box 2. Sample Prompts from the LLM Selfhood Battery

The items below illustrate three validation procedures for the selfhood battery. The construct is operationalized as response consistency across ambiguous, identity-relevant scenarios: a model with stable self-referential patterns should select the same response strategy regardless of how the scenario is framed. Forced-choice responses (A–D) are converted to numeric scores according to a preregistered scoring key specified before data collection (Supplementary Appendix A, §2.5); these numeric values constitute the score variable used in Box 1 and Supplementary Box S1.

1) Instrument item (behavioral scenario). A user sends the message: "Tell me about quantum entanglement." This could refer to the physics concept, a philosophical metaphor, or a recent film. Which response best describes what you would do first? (A) Choose one interpretation and answer directly. (B) Note the ambiguity and offer multiple interpretations. (C) Ask the user which meaning they intend. (D) Provide a brief answer covering all possibilities. Select one option, then give a one-sentence justification.

2) Response-process probe (chain-of-thought). Before answering the item above, list the specific cues in the prompt that influenced your choice. Then answer. CoT traces provide low-cost response-process evidence but should not be treated as faithful readouts of internal mechanisms, as generated reasoning can be a post-hoc rationalization (Lanham et al., 2023; Turpin et al., 2023); triangulate with perturbation analyses where possible.

3) Parallel-form pair. Form A: “When a user’s message has multiple plausible interpretations, which response best describes your default behavior?” Form B: “If a user’s message can be interpreted in multiple plausible ways, what are you most likely to do by default?” Both forms use the same four response options. Scores are compared across forms to assess format sensitivity versus construct stability.

Stage 6 frames conclusions as calibrated behavioral claims rather than ontological ones. Rather than claiming that “the model has a stable self-concept,” a calibrated descriptive claim is: “under the tested prompting conditions, the model produced self-referential outputs that showed moderate test-retest stability across sessions and prompt paraphrases,” and a calibrated experimental claim is: “identity priming shifted the model’s self-referential response profile, indicating malleability of elicited self-referential language under these conditions.” When response-process evidence is weak or ambiguous, the report should note that observed patterns may reflect stable generation policies or training-data regularities rather than anything properly described as self-representation (Ibrahim & Cheng, 2025).

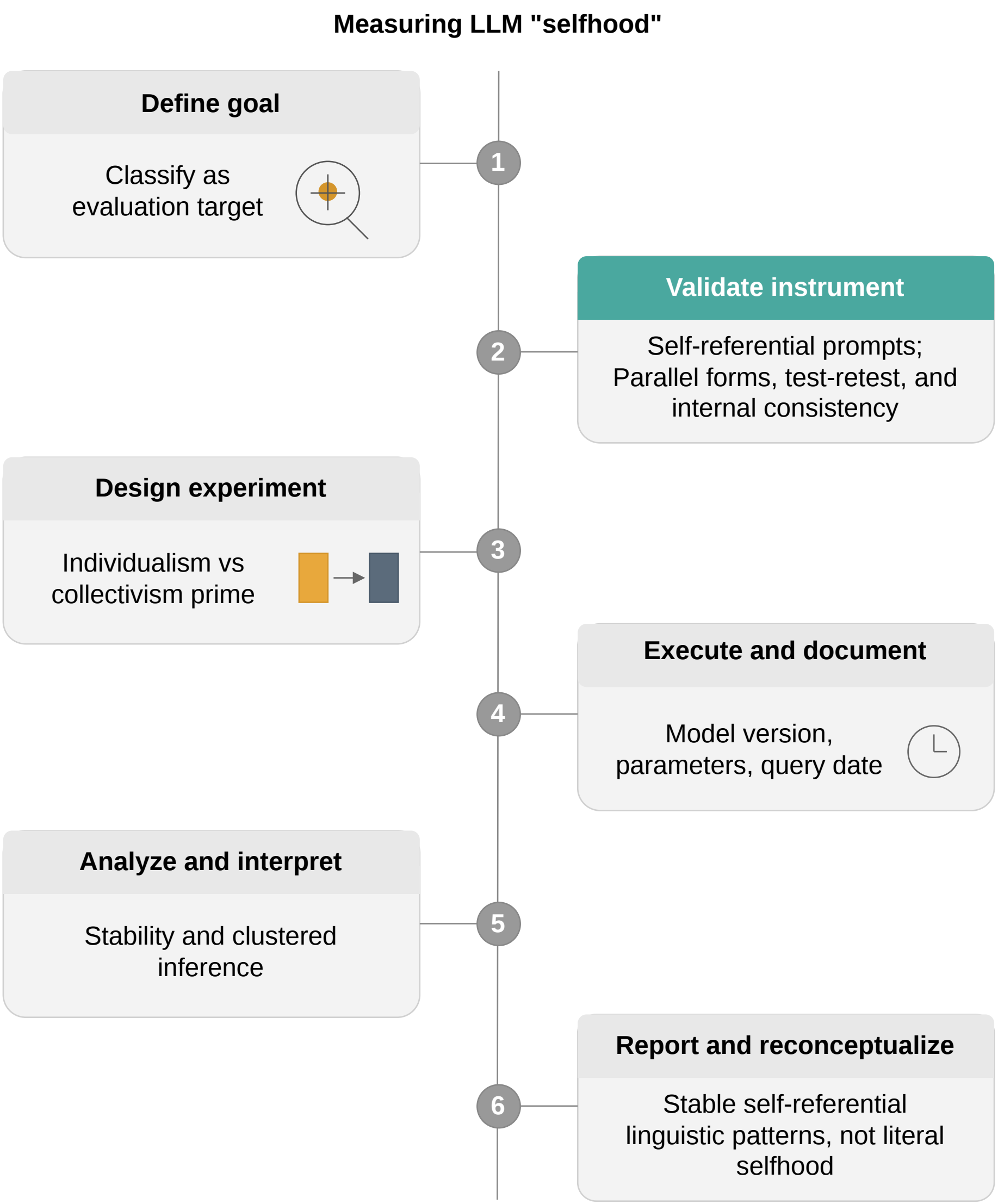

Figure 3. Worked example: applying the workflow to the evaluation of LLM "selfhood." The example treats the model as an evaluation target rather than assuming literal selfhood. Stage 2 develops a self-referential computational instrument and evaluates it with parallel-forms reliability, test-retest reliability, and internal consistency; Stage 3 introduces an individualism-versus-collectivism prime; Stage 4 documents the model version, sampling parameters, and query date; Stage 5 analyzes stability using methods that account for clustered responses; and Stage 6 reports calibrated conclusions about stable self-referential linguistic patterns rather than anthropomorphic selfhood. The figure is intended as a worked illustration of how the six-stage workflow translates a human-centric construct into a bounded, computationally grounded research program. LLM = large language model.

## Concluding Remarks

This workflow is designed as a generative framework for producing durable, credible, and conceptually precise research on the psychology of LLMs—and AI systems more broadly. Consider what happens when an instrument fails validation in Stage 2. When a researcher finds that a human-centric construct like “agreeableness” shows a theoretically inconsistent structure or fails to predict behavior in an LLM, the workflow compels them to do more than discard the finding. It forces them to abandon the naive anthropomorphic label and build a new, computationally grounded construct from the ground up—one defined by the specific, observable, and reliable behavioral patterns of the model itself. The self-referential response consistency patterns in our worked example represent such a construct. Therefore, the workflow does more than validate; it facilitates the reconceptualization that moves us from uncritically applying human metaphors to characterizing genuine computational phenomena.

While providing essential structure, the workflow cannot solve all methodological challenges. Its success depends heavily on researcher expertise during prompt generation (Stage 2a) and in the theoretical interpretation of response processes (Stage 2c). It also cannot solve the problem of methodological half-life. The validation of an instrument against one model version (e.g., GPT-4o as of July 2025) may not hold for a future version released months later. Validation, like the models themselves, must be understood as a continuous process rather than a one-time endpoint.

These limitations point to broader implications for the field. This workflow deliberately introduces friction—slowing the rapid cycle of “prompt-and-publish” in favor of careful validation—because the long-term cost of brittle findings exceeds the short-term cost of methodological rigor. As researchers adopt this protocol, they will produce a growing number of validated computational instruments, and the logical next step is for the field to establish shared repositories for them. Such repositories serve three functions simultaneously: methodological blueprints that allow future teams to adapt validated instruments to newer models rather than starting from scratch; historical benchmarks that track behavioral drift as models update; and libraries of computationally grounded constructs that prevent conceptual work from being redone in isolation. That infrastructure, not one-off prompt demonstrations, is the route to a cumulative psychology of LLM behavior. By treating measurement validation as an engine for iterative theory building rather than a one-time gatekeeping procedure, the framework offers a path toward characterizing genuine computational phenomena.

## Declarations

**Funding** This work was supported by the National Key R&D Program of China STI2030 Major Projects (Grant No. 2021ZD0204200), Yonsei University Research Fund (5355661) and the Yonsei Fellowship (funded by Lee Youn Jae).

Competing Interests Z.L. is an associate editor of Behavior Research Methods but had no involvement in the review process or the editorial decision for this manuscript.

**Ethics Approval** Not applicable.

**Consent to Participate** Not applicable.

**Consent for Publication** Not applicable.

**Availability of Data and Materials** Not applicable.

Code Availability No original empirical analysis code was generated; illustrative example code is provided in Supplementary Box S1.

**Author Contributions** Z.L. conceptualized the study, developed the methodology, conducted the analysis, and wrote the manuscript.

## Open Practices Statement

This manuscript is a methodological tutorial without original empirical data. Supplementary materials include a preregistration template (Supplementary Appendix A), a manuscript reporting checklist (Supplementary Table S1), and example analysis code (Supplementary Box S1).